%% file: main.tex
\documentclass[acmtog,screen,nonacm]{acmart}
\makeatletter
\@ACM@balancefalse
\makeatother
\setcopyright{none}
\acmDOI{}
\usepackage{booktabs}
\usepackage{graphicx}
\usepackage{amsmath}
\usepackage{array}
\usepackage{tabularx}
\usepackage{multirow}
\usepackage{url}
\usepackage[above,below]{placeins}
\usepackage{enumitem}
\author{Avinash Holenarasipur Nagendra}
\orcid{0009-0002-2800-5113}
\authornote{Avinash Holenarasipur Nagendra and Anand Kaniyaar Sriranga Prasad contributed equally and are joint first authors.}
\affiliation{%
  \department{Department of Heritage Science \& Technology}
  \institution{Indian Institute of Technology Hyderabad}
  \city{Sangareddy}
  \state{Telangana}
  \country{India}}
\email{ht23resch11003@iith.ac.in}

\author{Anand Kaniyaar Sriranga Prasad}
\orcid{0009-0008-1360-7334}
\authornotemark[1]
\affiliation{%
  \institution{HTSR Institute}
  \city{Bengaluru}
  \state{Karnataka}
  \country{India}}
\affiliation{%
  \department{Department of Heritage Science \& Technology}
  \institution{Indian Institute of Technology Hyderabad}
  \city{Sangareddy}
  \state{Telangana}
  \country{India}}

\author{Kousik Sarathy Sridharan}
\orcid{0000-0003-3736-1873}
\affiliation{%
  \department{Department of Heritage Science \& Technology}
  \institution{Indian Institute of Technology Hyderabad}
  \city{Sangareddy}
  \state{Telangana}
  \country{India}}
\affiliation{%
  \department{Department of Biomedical Engineering}
  \institution{Indian Institute of Technology Hyderabad}
  \city{Sangareddy}
  \state{Telangana}
  \country{India}}
\affiliation{%
  \institution{Bheri Technology Foundation}
  \streetaddress{ITIC, Indian Institute of Technology Hyderabad}
  \city{Sangareddy}
  \state{Telangana}
  \country{India}}

\author{Mohan Raghavan}
\orcid{0000-0002-9196-8193}
\affiliation{%
  \department{Department of Heritage Science \& Technology}
  \institution{Indian Institute of Technology Hyderabad}
  \city{Sangareddy}
  \state{Telangana}
  \country{India}}
\affiliation{%
  \department{Department of Biomedical Engineering}
  \institution{Indian Institute of Technology Hyderabad}
  \city{Sangareddy}
  \state{Telangana}
  \country{India}}
\affiliation{%
  \institution{Central Sanskrit University}
  \city{New Delhi}
  \country{India}}
\affiliation{%
  \institution{Bheri Technology Foundation}
  \streetaddress{ITIC, Indian Institute of Technology Hyderabad}
  \city{Sangareddy}
  \state{Telangana}
  \country{India}}

\input{author_layout}

\title{Dh\={a}r\={a} Bench: An Evidence-Oriented Retrieval Benchmark for Indian Knowledge Systems}

\begin{document}
\begin{abstract}
Dh\={a}r\={a} Bench is a retrieval benchmark for Indian Knowledge Systems (IKS), with 325 questions in 12 categories and 60,121 text chunks from 15 source groups. The categories were developed based on 4,500 questions submitted to the Dh\={a}r\={a} application; the evaluation set combines corpus-aligned application queries with expert-contributed questions. The annotations comprise 7,578 question-chunk relevance links involving 6,847 distinct corpus chunks. For 34 composite and multi-hop questions, 401 conjugate groups specify alternative sets of passages that jointly support each question. Chunk identifiers and hierarchical references connect the judgments to their source texts. The paper describes corpus preparation, annotation, and validation, and reports a fixed BM25 baseline implemented with Pyserini with $k_1=0.9$ and $b=0.4$. Across all questions, BM25 achieves MRR@20 of 0.315, Recall@10 of 0.182, and nDCG@10 of 0.205. On the group-annotated subset, complete-evidence scores are CE\_Recall@20 = 0.294 and CE\_MRR@20 = 0.141. The category results expose particularly weak verse retrieval, while group scores reveal gaps between finding individual passages and completing their support. The corpus and evaluation code are available at \url{https://github.com/Avinash2236/Dhara-Bench}.
\end{abstract}

\keywords{Indian Knowledge Systems, information retrieval, evidence retrieval, benchmark, provenance, question taxonomy}
\maketitle
\hypersetup{pdfcreator={LaTeX with hyperref}}

\input{sections/01_introduction}
\input{sections/02_related_work}
\input{sections/03_corpus}
\input{sections/04_methodology}
\input{sections/05_taxonomy}
\input{sections/06_ground_truth}
\input{sections/07_characterization}
\input{sections/08_evaluation}
\input{sections/09_results}
\input{sections/09_discussion}
\input{sections/10_conclusion}
\input{sections/acknowledgements}

\input{references_manual}
\end{document}

%% file: author_layout.tex
\makeatletter
\newcommand{\dharaauthoraffiliations}[2]{%
  \ifcsdef{typeset@author#1}{%
    \csletcs{dhara@typeset@author#1}{typeset@author#1}%
  }{%
    \cslet{dhara@typeset@author#1}{\@firstofone}%
  }%
  \csdef{typeset@author#1}##1{%
    \csuse{dhara@typeset@author#1}{##1}\textsuperscript{#2}}%
}
\dharaauthoraffiliations{1}{1}
\dharaauthoraffiliations{2}{2,1}
\dharaauthoraffiliations{3}{1,3,5}
\dharaauthoraffiliations{4}{1,3,4,5}

\def\dhara@skipaffiliation#1#2{%
  \let\and\@typeset@author@line}
\patchcmd{\@mkauthors@i}
  {\noindent\addresses\@typeset@author@line}
  {\noindent\let\affiliation\dhara@skipaffiliation\addresses\@typeset@author@line}
  {}%
  {\ClassError{dhara-paper}{Cannot apply shared-affiliation layout}%
    {Check the author layout against the installed acmart class.}}
\makeatother

\authorsaddresses{%
  Affiliations: \textsuperscript{1}Department of Heritage Science \& Technology,
  Indian Institute of Technology Hyderabad, Sangareddy, Telangana, India.\newline
  \textsuperscript{2}HTSR Institute, Bengaluru, Karnataka, India.\newline
  \textsuperscript{3}Department of Biomedical Engineering,
  Indian Institute of Technology Hyderabad, Sangareddy, Telangana, India.\newline
  \textsuperscript{4}Central Sanskrit University, New Delhi, India.\newline
  \textsuperscript{5}Bheri Technology Foundation, ITIC, Indian Institute of Technology Hyderabad, Sangareddy, Telangana, India.\newline
  Corresponding author: Avinash Holenarasipur Nagendra,
  \href{mailto:ht23resch11003@iith.ac.in}{\nolinkurl{ht23resch11003@iith.ac.in}}.}

%% file: sections/01_introduction.tex
\section{Introduction}
India's intangible heritage includes knowledge sustained through oral traditions, written texts, formal learning, and community practice. The term Indian Knowledge Systems (IKS) covers these codified traditions and the practical expertise passed down within communities~\cite{kapoor2005iks}. Its fields include philosophy, grammar, logic, mathematics, astronomy, architecture, medicine, agriculture, metallurgy, and trade, alongside literature and everyday practices such as dress, language, ethics, and cuisine~\cite{raghavan2023elements}.

Indian philosophical inquiry gives attention to both the object of knowledge (\emph{prameya}) and the means of valid knowledge (\emph{pram\={a}\d{n}a})~\cite{phillips2024epistemology}. Yoga accepts three means: perception (\emph{pratyak\d{s}a}), inference (\emph{anum\={a}na}), and authoritative testimony (\emph{\={a}gama})~\cite{bryantYoga}. Advaita Ved\={a}nta recognizes six: perception, inference, testimony (\emph{\'{s}abda}), comparison (\emph{upam\={a}na}), postulation (\emph{arth\={a}patti}), and non-cognition (\emph{anupalabdhi})~\cite{srinivasAdvaita}. These traditions combine observation, reasoning, and transmitted knowledge. Working with their texts therefore calls for examining the origin, relevance, and validity of supporting material before drawing conclusions.

Finding that evidence is a substantial task. A reader may seek a verse from its opening words, the meaning of an Indic term, a sequence of events, or passages supporting an interpretation. Prose chapters, verses, s\={u}tras, and commentaries also differ in length and reference conventions. Some questions can be addressed from one passage; others require material from several locations.

Dh\={a}r\={a} Bench provides a shared corpus and expert annotations for studying this retrieval problem. A system ranks existing text chunks in response to a question, and evaluation compares their identifiers with the annotated passages. Here, evidence means inspectable textual support. The validity of source claims and the correctness of answers synthesized from them require separate assessment.

The collection contains 325 questions in 12 categories and 60,121 chunks from 15 source groups. Questions were selected from corpus-aligned Dh\={a}r\={a} application queries and supplemented by domain experts. The annotations record 7,578 question-chunk links covering 6,847 distinct chunks. For 19 composite and 15 multi-hop questions, 401 \emph{conjugate evidence groups} identify alternative sets of passages that jointly support the whole question. These groups allow evaluation to distinguish finding a useful passage from assembling sufficient support.

The paper describes source preparation, question selection, annotation, and validation. A fixed BM25 baseline supplies overall and category-wise results, including scores for group completion. The accompanying GitHub release provides the corpus, questions, annotations, source metadata, rankings, and evaluation code.

%% file: sections/02_related_work.tex
\section{Related Work}
Table~\ref{tab:related} places Dh\={a}r\={a} Bench alongside retrieval, Indian-language, cultural, and IKS resources. The principal differences concern the texts being searched and the units used for evaluation.

\begin{table*}[!t]
\caption{Positioning relative to representative retrieval, Indian-language, cultural, and IKS-related resources.}
\label{tab:related}
\begin{tabularx}{\textwidth}{>{\raggedright\arraybackslash}p{0.14\textwidth} >{\raggedright\arraybackslash}p{0.20\textwidth} >{\raggedright\arraybackslash}p{0.21\textwidth} >{\raggedright\arraybackslash}p{0.20\textwidth} >{\raggedright\arraybackslash}X}
\toprule
Resource & Primary objective & Knowledge scope & Evidence / grading unit & Relation to Dh\={a}r\={a} Bench \\
\midrule
BEIR~\cite{thakur2021beir} & Information retrieval & Diverse domains; no dedicated IKS collection & Dataset-specific relevance judgments & General heterogeneous IR benchmark \\
KILT~\cite{petroni2021kilt} & Knowledge-intensive NLP & Wikipedia & Provenance and task output & Provenance-aware multi-task benchmark \\
Anveshana~\cite{jagadeeshan2025anveshana} & Cross-lingual retrieval & \emph{\'{S}r\={i}madbh\={a}gavatam} & English query-Sanskrit document pairs & Retrieval within one IKS work \\
VedantaNY-10M~\cite{mandikal2024ancient} & RAG / long-form QA & Advaita Ved\={a}nta & Retrieved datastore passages & Specialized philosophy retrieval setting \\
Indic-QA~\cite{singh2025indicqa} & Context-grounded QA & 11 Indian languages & Supplied QA context & Multilingual QA \\
SANSKRITI~\cite{maji2025sanskriti} & Cultural knowledge & Indian culture & QA-pair answer evaluation & Cultural knowledge evaluation \\
IndQA~\cite{openai2025indqa} & Cultural reasoning & Indian languages and cultural domains & Expert rubric and ideal answer & Expert-authored reasoning benchmark \\
IKS-Instruct~\cite{singaravelu2026iks} & Instruction tuning & Multilingual IKS material & Instruction-response pairs & IKS adaptation resource \\
Dh\={a}r\={a} Bench & Evidence retrieval & 15 IKS source groups & Chunk relevance and joint evidence groups & 12 question categories over a fixed IKS corpus \\
\bottomrule
\end{tabularx}
\end{table*}

\paragraph{Retrieval and provenance benchmarks.}
BEIR's original suite of 18 datasets covers domains such as biomedical literature, finance, news, and Wikipedia, but has no dedicated IKS collection~\cite{thakur2021beir}. KILT uses a shared Wikipedia snapshot to evaluate knowledge-intensive NLP tasks and their provenance~\cite{petroni2021kilt}. Encyclopedic coverage can include IKS topics; Dh\={a}r\={a} Bench instead organizes retrieval judgments around the selected source works and their internal textual locations.

\paragraph{Indian-language and cultural evaluation.}
Indic-QA covers extractive and abstractive question answering in 11 Indian languages~\cite{singh2025indicqa}. SANSKRITI assesses Indian cultural knowledge~\cite{maji2025sanskriti}, and IndQA uses expert-authored questions and rubrics to assess culturally grounded knowledge and reasoning~\cite{openai2025indqa}. These resources overlap with IKS in language or subject matter and evaluate answers. Dh\={a}r\={a} Bench supplies ranked-passage judgments over a fixed corpus.

\paragraph{Sanskrit and IKS resources.}
Anveshana pairs 3,400 English queries with Sanskrit documents from \emph{\'{S}r\={i}madbh\={a}gavatam} for cross-lingual retrieval~\cite{jagadeeshan2025anveshana}. Mandikal~\cite{mandikal2024ancient} studies retrieval-augmented long-form question answering in Advaita Ved\={a}nta using the VedantaNY-10M datastore of public discourses. SanskritShala offers Sanskrit NLP tools for teaching and annotation~\cite{sandhan2023sanskritshala}, while IKS-Instruct supplies multilingual instruction-response data for model adaptation~\cite{singaravelu2026iks}. The contribution here is a common retrieval evaluation across 15 source groups, 12 question categories, and expert-selected passages.

\paragraph{Joint supporting evidence.}
HotpotQA annotates sentence-level supporting facts across documents~\cite{yang2018hotpotqa}; MuSiQue connects questions through dependent reasoning steps~\cite{trivedi2022musique}; and Multi-Hop Dense Retrieval studies supporting-passage retrieval for complex questions~\cite{xiong2021mdr}. Dh\={a}r\={a} Bench records alternative chunk groups for composite and multi-hop questions. Section~\ref{sec:ground-truth} defines this representation, and Section~\ref{sec:ce-metrics} specifies how group completion is scored.

%% file: sections/03_corpus.tex
\section{Source Corpus}
\subsection{Selection of works}
The 15 source groups cover epic narratives, Pur\={a}\d{n}ic literature, Vedic hymns, philosophical dialogues, s\={u}tras, and story collections, together with treatises on medicine, performing arts, and architecture. Modern studies cover mathematics and South Indian history. Table~\ref{tab:sources} lists the works and their principal IKS domains.

Selection sought variety in textual form and subject matter. Some groups contain several editions, translations, or commentaries on one work. The Bhagavadg\={i}t\={a} forms a separate group because the corpus includes dedicated translation and commentary resources for it.

\begin{table}[!htbp]
\caption{The 15 source groups in Dh\={a}r\={a} Bench and their principal IKS domains.}
\label{tab:sources}
\begin{tabularx}{\linewidth}{r>{\raggedright\arraybackslash}X>{\raggedright\arraybackslash}p{0.24\linewidth}}
\toprule
No. & Source group & IKS domain \\
\midrule
1 & Mah\={a}bh\={a}rata & Itih\={a}sa \\
2 & V\={a}lm\={i}ki R\={a}m\={a}ya\d{n}a & Itih\={a}sa \\
3 & Agni Pur\={a}\d{n}a & Pur\={a}\d{n}a \\
4 & \={A}pastamba Dharma-s\={u}tra & Dharma\'{s}\={a}stra \\
5 & Caraka Sa\d{m}hit\={a} & Ayurveda \\
6 & Kath\={a}sarits\={a}gara & K\={a}vya \\
7 & N\={a}\d{t}ya\'{s}\={a}stra & Kal\={a} \\
8 & A\d{s}\d{t}\={a}\.{n}ga H\d{r}dayam & Ayurveda \\
9 & Bhagavadg\={i}t\={a} & Philosophy \\
10 & Ka\d{t}ha Upani\d{s}ad & Upani\d{s}ad \\
11 & M\={a}nas\={a}ra & \'{S}ilpa \\
12 & History of Hindu Mathematics & Mathematics \\
13 & A History of South India & History \\
14 & \d{R}g \'{S}\={a}kala Sa\d{m}hit\={a} & Veda \\
15 & Yoga-s\={u}tra & Dar\'{s}ana \\
\bottomrule
\end{tabularx}
\end{table}

The corpus provides English translations and explanations, while retaining Sanskrit terminology, source expressions, and quotations where present. Machine translations were produced where required using a locally hosted, instruction-tuned Gemma 4 31B model (\texttt{gemma-4-31B-it})~\cite{gemmateam2026gemma4}. Domain-specific IKS technical terms are also retained in the questions.

\subsection{Acquisition and provenance}
Machine-readable web pages were acquired by text extraction; scanned PDFs were converted with Surya OCR~\cite{paruchuri2025surya}. The source manifest gives bibliographic descriptions and URLs for the resources within each group.

Each corpus record has 3 fields: \texttt{id}, its unique integer identifier; \texttt{chunk}, the searchable text; and \texttt{ref}, the hierarchical location in the source. Evaluation uses the identifier to connect a judgment to a specific passage; the reference and manifest allow that passage to be inspected in its textual and bibliographic context.

\subsection{Data and code availability}
\label{sec:data-availability}
The GitHub repository at \url{https://github.com/Avinash2236/Dhara-Bench} contains the following data files:
\begin{itemize}
\item \texttt{chunks.json}: all corpus chunk records.
\item \texttt{Dhara Bench eval set.json}: questions, category labels, gold chunk identifiers, and conjugate evidence groups.
\item \texttt{source\_metadata.csv}: source descriptions and URLs.
\end{itemize}
The repository also contains BM25 scripts in \texttt{scripts/}, and saved rankings and metric outputs in \texttt{metric\_outputs/}. Section~\ref{sec:ground-truth} defines the annotation fields, and Section~\ref{sec:reproducibility} explains how to reproduce the scores.

%% file: sections/04_methodology.tex
\section{Dataset Construction}\label{sec:construction}
\subsection{Question collection and taxonomy development}
\label{sec:question-collection}
Analysis of 4,500 questions submitted to the Dh\={a}r\={a} application\footnote{Dh\={a}r\={a} application: \url{https://dhara.bheri.in/}.} led to 12 question categories, T01-T12. Questions were grouped by their main requested operation: the same source could be queried for a fact, a list, a word meaning, a verse, or an interpretation. This produced an empirical scheme based on application use.

Most categories describe tasks found across domains, including fact lookup, enumeration, counting, chronology, composite queries, multi-hop reasoning, and false-premise correction. Lexical understanding (T05) and verse or textual retrieval (T08) are particularly relevant to scholarly discourse in IKS. T05 covers the meaning or equivalent of an Indic term, including its use within a verse. T08 covers locating, identifying, or completing a text from opening words, a numbered reference, or a paraphrased meaning. Such tasks also occur in other literary traditions; their distinctive demands here arise from Indic terminology, textual forms, and reference conventions.

Interpretive questions (T11) depend on concepts within a textual tradition, while computational questions (T12) draw on its rules, quantities, or procedures. T12 includes rule-based lexical derivation, distinguished from meaning lookup in T05. Table~\ref{tab:taxonomy} gives the full definitions.

The final 325 questions combine corpus-aligned application queries with questions supplied by domain experts. Each record retains its original wording, receives a unique \texttt{query\_id}, and has one primary category. The category counts describe this selected set, rather than the distribution of all 4,500 application questions.

\subsection{Chunk construction}
\label{sec:chunk-construction}
Prose chapters were segmented independently with LangChain's RecursiveCharacterTextSplitter~\cite{langchainRecursiveSplitter}, using a target of 1,800 characters and an overlap setting of 250 characters, measured by Python's \texttt{len}. Splits followed paragraph boundaries where possible, then line boundaries, periods followed by a space, and individual word separators. Chunks and overlaps stayed within chapter boundaries. Leading periods introduced by splitting and surrounding whitespace were removed.

A verse and its explicitly linked translation or commentary were kept together when they fit within approximately 1,800 characters. Longer units were split with their textual location retained. These rules account for records that exceed the prose target.

\subsection{Evidence annotation}
\label{sec:evidence-annotation}
Domain experts identified candidate passages using their knowledge of the source works. They were permitted to use the Dh\={a}r\={a} application and tools such as Gemini, ChatGPT, and Claude to supplement their expertise and search as comprehensively as possible for relevant evidence. Experts inspected candidate passages against corpus records, judged their relevance, and recorded the selected chunk identifiers. On average, 2 experts annotated each question. Their selected chunk identifiers were combined by taking the union of their annotations and removing duplicates.

Selection followed the question's task. A lexical question calls for an explanation or equivalent of the term; a verse query calls for the requested text or its location; and a computational question calls for the relevant premises or rules. For false-premise questions, experts selected passages that contradicted the assumption or established the correct fact. A missing passage alone was insufficient grounds for a correction.

The 19 Composite (T06) and 15 Multi-hop (T07) questions received group annotations as well. Experts specified the intended answer scope, separated it into required facts or relations, and inspected which requirements each candidate supported. T06 groups had to cover every part of the query; T07 groups had to supply the facts and connections in the reasoning chain. Entity identities, aliases, chronology, and compatibility between passages were considered when combining chunks.

Alternative combinations were retained when they supplied the same required support. Redundant supersets were removed, and a singleton was accepted when one chunk covered the whole scope. Section~\ref{sec:ground-truth} formalizes these groups and illustrates them with a released question. In the released annotations, the flat gold list for T06/T07 is the union of their groups; the other 10 categories use their reviewed passage lists.

%% file: sections/05_taxonomy.tex
\section{Question Taxonomy}
Each question receives the category of its main requested operation, including when it involves subsidiary steps. Table~\ref{tab:taxonomy} defines T01-T12 and gives their counts. The scheme emerged from the application-query analysis in Section~\ref{sec:question-collection}; Composite (T06) refers to multiple parts within a single query.

\begin{table*}[!t]
\caption{Question categories in Dh\={a}r\={a} Bench (325 questions).}
\label{tab:taxonomy}
\begin{tabularx}{\textwidth}{p{0.06\textwidth} p{0.25\textwidth} X r}
\toprule
Code & Question category & Task description & $n$ \\
\midrule
T01 & Fact retrieval & Explicitly stated information & 51 \\
T02 & List / set retrieval & Entities satisfying a condition & 40 \\
T03 & Counting / aggregation & Counts, totals, or numerical aggregation & 22 \\
T04 & Chronology / sequence & Order of events or related entities & 25 \\
T05 & Lexical understanding & Meaning, translation, or equivalence & 31 \\
T06 & Composite & Two or more independent questions grouped in one query & 19 \\
T07 & Multi-hop reasoning & Chains of linked facts across entities or events & 15 \\
T08 & Verse / textual retrieval & Retrieve, locate, identify, or complete a verse or text & 26 \\
T09 & False-premise correction & Identify and correct an incorrect assumption using source evidence & 18 \\
T10 & Summarization / expansion / composition & Structured overview, detailed exposition, or rule-constrained composition & 29 \\
T11 & Analytical / interpretive reasoning & Interpretation, abstraction, or thematic reasoning grounded in the textual tradition & 32 \\
T12 & Mathematical / logical / computational problems & Sequential computation, inference, algorithms, or rule-based lexical derivation & 17 \\
\midrule
 & \textbf{Total} & & \textbf{325} \\
\bottomrule
\end{tabularx}
\end{table*}

%% file: sections/06_ground_truth.tex
\section{Ground-Truth Evidence}
\label{sec:ground-truth}
Question 157 asks, ``What did Nachiketa ask for as his first boon? How did Yama respond?'' Answering the whole question requires evidence for both the request and the response. In Ka\d{t}ha Upani\d{s}ad I.1.10-11, chunks \texttt{262130} and \texttt{262131} contain the request and reply; chunks \texttt{261886} and \texttt{261887} contain their respective commentaries. The annotations record 4 alternatives, each pairing one request passage with one response passage. Retrieving only the request leaves the second part unsupported.

For each question $q$, the gold set $G_q$ lists the identifiers of chunks selected by domain experts as relevant to the requested task. Each selected chunk receives a binary relevance label, and its \texttt{ref} field locates the passage within its source. Section~\ref{sec:evidence-annotation} describes the selection procedure.

Composite (T06) and Multi-hop (T07) questions additionally have \emph{conjugate evidence groups}: combinations of passages judged jointly sufficient for the specified answer scope. Their collection is written as
\begin{equation}
\mathcal{C}_q=\{C_{q,1},\ldots,C_{q,m_q}\}.
\end{equation}
All chunks within a group are required (AND), while any one complete group provides an acceptable alternative (OR). Each group is minimal: removing any member leaves some required support missing. A group may contain a single chunk when that passage supports the whole question.

The JSON stores the groups in \texttt{refs.conjugate\_groups}. For T06/T07, \texttt{refs.chunk} contains the distinct chunk identifiers across all groups,
\begin{equation}
G_q=\bigcup_{C\in\mathcal{C}_q} C.
\end{equation}
For Question 157, this union contains 4 identifiers, although any one of the 4 recorded request-response pairs supplies the required support. Conventional metrics score retrieval of individual chunks from $G_q$; the group metrics in Section~\ref{sec:ce-metrics} test whether at least one complete group is retrieved.

For the remaining 10 categories, annotations list relevant chunks individually, without specifying which combinations jointly provide sufficient support. An individual relevance label does not imply that the chunk alone answers the whole question.

The annotations reflect evidence identified and reviewed by domain experts through their knowledge of the sources and tool-assisted searches. However, we cannot claim exhaustive coverage of every relevant passage or possible combination of supporting passages. Scores measure retrieval against the recorded judgments.

%% file: sections/07_characterization.tex
\section{Dataset Characterization}
\subsection{Corpus and evidence scale}
Table~\ref{tab:corpus-summary} reports corpus scale and annotation coverage. Most chunks fit the prose segmentation target, with a small tail of longer records.

The annotations involve 11.4\% of the corpus. A chunk used by several questions contributes once to the distinct count and once per question to the link count. T06/T07 statistics use each question's group union, counting a chunk once regardless of repeated membership across alternatives.

\begin{table}[!htbp]
\caption{Corpus and relevance-link summary for Dh\={a}r\={a} Bench.}
\label{tab:corpus-summary}
\centering
\begin{tabular}{lr}
\toprule
Statistic & Value \\
\midrule
Corpus chunks & 60,121 \\
Unique corpus IDs & 60,121 \\
Total Unicode characters & 50,655,188 \\
Whitespace-delimited words & 8,450,776 \\
Median chunk length (characters) & 615 \\
90th-percentile chunk length & 1,723 \\
Chunks $\leq$ 1,800 characters & 59,445 (98.88\%) \\
Question-chunk relevance links & 7,578 \\
Distinct gold chunks & 6,847 (11.4\% of corpus) \\
Orphan gold IDs & 0 \\
Duplicate corpus IDs & 0 \\
Empty chunks / missing refs & 0 / 0 \\
\bottomrule
\end{tabular}
\end{table}

\begin{figure*}[!t]
\centering
\includegraphics[width=0.95\textwidth]{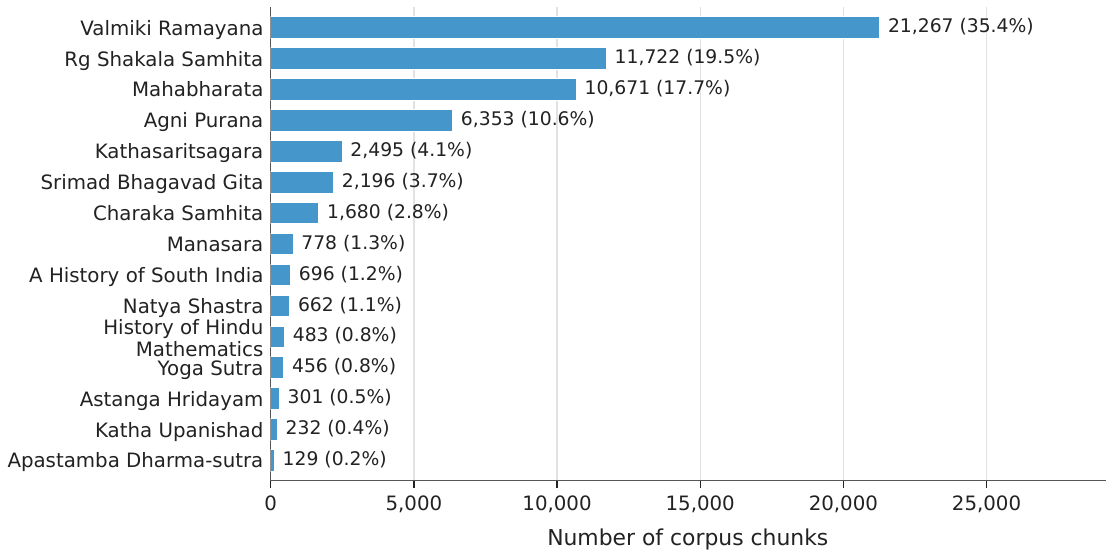}
\caption{Corpus chunks by source group. Labels give the chunk count and percentage of the full corpus.}
\label{fig:chunks-source}
\Description{Bar chart of chunk counts for the 15 source groups. Valmiki Ramayana is the largest group with 21,267 chunks, followed by Rg Shakala Samhita with 11,722 and Mahabharata with 10,671. Each bar is labeled with its count and percentage of the full corpus.}
\end{figure*}

Figure~\ref{fig:chunks-source} shows an uneven distribution of chunks across source groups. Such variation is expected in an IKS corpus, where expansive epics and textual collections coexist with much shorter works. Differences in textual extent partly explain why a few sources contribute a large share of the corpus: the R\={a}m\={a}ya\d{n}a, \d{R}g \'{S}\={a}kala Sa\d{m}hit\={a}, and Mah\={a}bh\={a}rata together account for 72.6\% of all chunks. This imbalance should be considered when interpreting retrieval performance across the corpus.

\begin{figure*}[!t]
\centering
\includegraphics[width=0.95\textwidth]{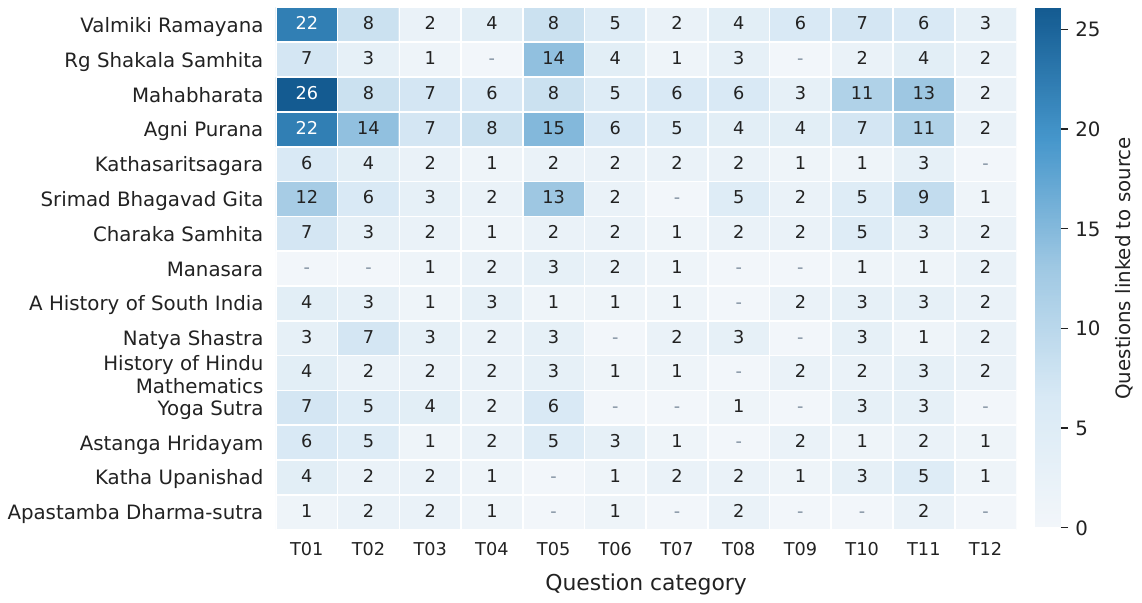}
\caption{Question-source coverage by question category. Rows follow the source-size order in Figure~\ref{fig:chunks-source}; cells count questions, and dashes indicate zero. A question linked to multiple sources contributes once to each source group.}
\label{fig:cat-source}
\Description{Blue sequential heatmap with source groups on rows and question categories T01 through T12 on columns. Cell values give the number of questions linked to each source; dashes indicate zero. Darker blue means more questions. A question can contribute to multiple sources.}
\end{figure*}

Figure~\ref{fig:cat-source} maps the annotated questions to their sources. The mix differs by category, so a category comparison also reflects which works supply its passages. Cell counts include each source used by a question.

\subsection{Chunk-length distribution}
\begin{figure}[!htbp]
\centering
\includegraphics[width=\linewidth]{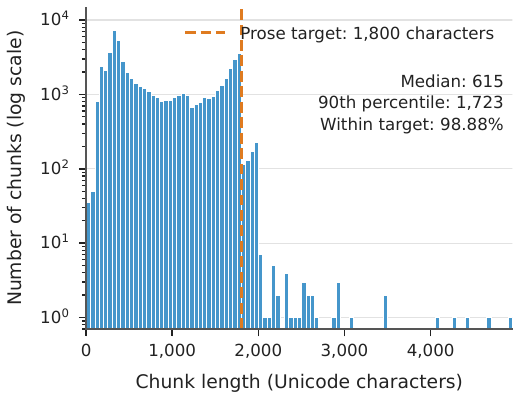}
\caption{Distribution of corpus chunk lengths in Unicode characters. The dashed line marks the 1,800-character target used for prose segmentation.}
\label{fig:chunk-length}
\Description{Histogram of corpus chunk lengths in Unicode characters. The median is 615 and the 90th percentile is 1,723. A dashed vertical line marks 1,800 characters; 98.88 percent of chunks are at or below this target.}
\end{figure}

Figure~\ref{fig:chunk-length} shows the full length distribution. Only 676 chunks (1.12\%) exceed the prose target, and the maximum is 4,920 characters. Chunks containing verses and their associated translations or commentaries were formed using the segmentation rules described in Section~\ref{sec:chunk-construction}.

\subsection{Evidence-set variation}
Table~\ref{tab:evidence-size} shows substantial differences in annotation size. T10 has a median of 21 chunks, a mean of 149.38, and the largest set of 2,549. T08 and T12 have medians of 1.5 and 2. A fixed retrieval budget therefore covers very different fractions of the annotated material across categories.

\begin{table}[!htbp]
\caption{Gold-evidence set size by question category.}
\label{tab:evidence-size}
\centering
\begin{tabular}{lrrrr}
\toprule
Category & $n$ & Median & Mean & Max \\
\midrule
T01 & 51 & 9.0 & 12.53 & 49 \\
T02 & 40 & 11.0 & 11.93 & 41 \\
T03 & 22 & 9.0 & 11.82 & 64 \\
T04 & 25 & 10.0 & 24.92 & 144 \\
T05 & 31 & 8.0 & 9.48 & 29 \\
T06 & 19 & 4.0 & 7.16 & 31 \\
T07 & 15 & 4.0 & 5.07 & 16 \\
T08 & 26 & 1.5 & 4.50 & 29 \\
T09 & 18 & 5.0 & 6.00 & 17 \\
T10 & 29 & 21.0 & 149.38 & 2,549 \\
T11 & 32 & 11.0 & 11.94 & 40 \\
T12 & 17 & 2.0 & 7.88 & 44 \\
\midrule
Overall & 325 & 7.0 & 23.32 & 2,549 \\
\bottomrule
\end{tabular}
\end{table}

\begin{figure}[!htbp]
\centering
\includegraphics[width=\linewidth]{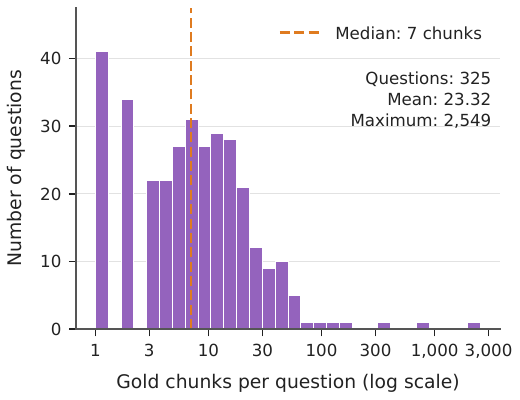}
\caption{Distribution of deduplicated gold chunks per question on a logarithmic horizontal axis. The dashed line marks the median of 7 chunks.}
\label{fig:gold-dist}
\Description{Histogram of the number of gold chunks per question, with a logarithmic horizontal axis. The distribution is right-skewed, with median 7, mean 23.32, and maximum 2,549 chunks.}
\end{figure}

\begin{figure}[!htbp]
\centering
\includegraphics[width=\linewidth]{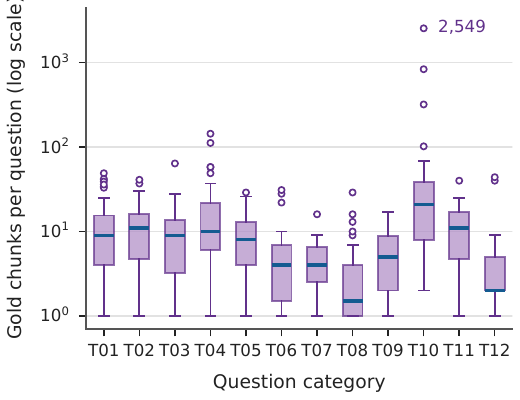}
\caption{Gold-evidence set size by question category on a logarithmic vertical axis. Boxes span the interquartile range, blue lines mark medians, and whiskers extend to the most extreme observations within 1.5 interquartile ranges of the box. Circles show observations beyond the whiskers.}
\label{fig:gold-cat}
\Description{Comparison of gold-set sizes for categories T01 through T12 on a logarithmic vertical scale. T10 has median 21 and maximum 2,549; T08 and T12 have medians 1.5 and 2. Table~\ref{tab:evidence-size} gives the category statistics.}
\end{figure}

Figures~\ref{fig:gold-dist} and~\ref{fig:gold-cat} show the long tail of gold-set sizes and its distribution across categories. Logarithmic axes retain the extreme observations while keeping the smaller sets visible.

The group annotations comprise 330 T06 groups and 71 T07 groups. Of the 401 total, 35 are singletons, 340 contain 2 chunks, 24 contain 3, and 2 contain 4; 366 therefore require multiple chunks. Their per-question unions contribute 212 relevance links: 136 for T06 and 76 for T07. These group counts describe alternative support combinations, whereas Table~\ref{tab:evidence-size} counts distinct chunks per question.

%% file: sections/08_evaluation.tex
\section{Evaluation Protocol and BM25 Baseline}
\label{sec:eval}
\subsection{Metrics}
\label{sec:metrics}
Individual-chunk scores comprise Precision, Recall, F1, mean reciprocal rank (MRR), and normalized discounted cumulative gain (nDCG). Group-completion scores additionally use the T06/T07 annotations. Let $Q$ be the 325 questions, $G_q$ the nonempty gold set from Section~\ref{sec:ground-truth}, and $L_q=(d_{q,1},\ldots,d_{q,n_q})$ the deduplicated ranking. For cutoff $k$, $L_q^{(k)}$ contains the first $\min(k,n_q)$ identifiers and $E_q(k)=L_q^{(k)}\cap G_q$ contains the gold hits.

Precision, Recall, F1, and nDCG are computed at $k\in\{3,8,10,20\}$. For any of these per-question measures $M@k(q)$, the overall score is the macro-average
\begin{equation}
\overline{M@k}=\frac{1}{|Q|}\sum_{q\in Q}M@k(q).
\end{equation}
Each question has equal weight. Category scores take the same mean over their subset of questions. All scores range from 0 to 1, with higher values indicating better agreement with the judgments; chunks outside $G_q$ receive no relevance credit.

\subsubsection{Precision at $k$}
Precision@$k$ measures the proportion of the first $k$ retrieval positions occupied by gold chunks~\cite{manning2008ir}:
\begin{equation}
P@k(q)=\frac{|E_q(k)|}{k}.
\end{equation}
The denominator remains $k$ for shorter result lists, so unfilled positions receive no credit.

\subsubsection{Recall at $k$}
Recall@$k$ measures the fraction of a question's gold evidence recovered within the first $k$ results~\cite{manning2008ir}:
\begin{equation}
R@k(q)=\frac{|E_q(k)|}{|G_q|}.
\end{equation}
The maximum at a fixed cutoff is $\min(k,|G_q|)/|G_q|$: full recall requires enough result positions to hold the gold set.

\subsubsection{F1 at $k$}
F1@$k$ combines precision and recall through their harmonic mean~\cite{manning2008ir}:
\begin{equation}
F1@k(q)=
\begin{cases}
\displaystyle\frac{2P@k(q)R@k(q)}{P@k(q)+R@k(q)}, & P@k(q)+R@k(q)>0,\\[6pt]
0, & \text{otherwise}.
\end{cases}
\end{equation}
F1 is calculated per question before averaging. This macro-average generally differs from the harmonic mean of macro Precision and Recall.

\subsubsection{Mean reciprocal rank at 20}
Mean reciprocal rank summarizes how early the first relevant result appears~\cite{voorhees1999qa}. We report MRR@20, which credits the first gold hit only when it occurs within the first 20 results. Let $r_q$ be the rank of the first gold chunk in the returned list, or $\infty$ if none is retrieved. The per-question reciprocal rank at this cutoff and its mean are
\begin{align}
\mathrm{RR}@20(q)&=
\begin{cases}
1/r_q, & r_q\leq 20,\\
0, & \text{otherwise},
\end{cases}\\
\mathrm{MRR}@20&=\frac{1}{|Q|}\sum_{q\in Q}\mathrm{RR}@20(q).
\end{align}
A first gold hit at rank 1 contributes 1; at rank 5 it contributes 0.2. Only the rank of the first relevant chunk contributes to MRR; additional relevant chunks do not increase the score.

\subsubsection{Normalized discounted cumulative gain at $k$}
nDCG@$k$ evaluates the placement of relevant chunks throughout the first $k$ results, assigning greater weight to earlier ranks~\cite{jarvelin2000ir,manning2008ir}. We use binary relevance: $\mathrm{rel}_{q,i}=1$ when the chunk at rank $i$ belongs to $G_q$, and zero otherwise, including unfilled positions. Discounted cumulative gain (DCG), ideal DCG, and nDCG are defined as
\begin{align}
\mathrm{DCG}@k(q)&=\sum_{i=1}^{k}\frac{\mathrm{rel}_{q,i}}{\log_2(i+1)},\\
\mathrm{IDCG}@k(q)&=\sum_{i=1}^{\min(k,|G_q|)}\frac{1}{\log_2(i+1)},\\
\mathrm{nDCG}@k(q)&=\frac{\mathrm{DCG}@k(q)}{\mathrm{IDCG}@k(q)}.
\end{align}
IDCG places gold chunks first, up to the cutoff. It is positive because every gold set is nonempty. Unlike MRR, nDCG rewards the positions of multiple relevant results.

\subsubsection{Complete-Evidence Recall (CE\_Recall) at $k$}
\label{sec:ce-metrics}
For T06 and T07, let $Q_{\mathrm{CE}}=\{q:\mathcal{C}_q\neq\varnothing\}$ contain the questions with at least one conjugate group. Complete-Evidence Recall (CE\_Recall) at cutoff $k$ is
\begin{equation}
\mathrm{CE\_Recall}@k(q)=
\begin{cases}
1, & \exists C\in\mathcal{C}_q:\ C\subseteq L_q^{(k)},\\
0, & \text{otherwise}.
\end{cases}
\end{equation}
This binary score records whether the ranking contains a whole group. Its macro-average,
\begin{equation}
\mathrm{CE\_Recall}@k=\frac{1}{|Q_{\mathrm{CE}}|}\sum_{q\in Q_{\mathrm{CE}}}\mathrm{CE\_Recall}@k(q),
\end{equation}
is the fraction of questions with a completed group. It is reported at $k\in\{3,8,10,20\}$. The requirement to recover a complete evidence set has precedents in FEVER~\cite{thorne2018fever} and KILT~\cite{petroni2021kilt}. CE\_Recall counts questions for which at least one annotated group is complete; KILT's Recall@$k$ instead measures the fraction of provenance sets recovered under its ranking convention. All 34 annotated questions are evaluable. A question lacking a valid group would be excluded and counted as unevaluable; individual-chunk scores cover all 325 questions.

\subsubsection{Complete-Evidence reciprocal rank and mean}
We adapt reciprocal rank~\cite{voorhees1999qa} to the earliest rank at which a complete evidence group is available. KILT similarly uses the lowest-ranked required page to locate a provenance set~\cite{petroni2021kilt}; our definition retains the original chunk positions and takes the earliest completion across alternative groups. We use CE\_RR for the per-question score and CE\_MRR for its mean over questions.

Let $\operatorname{rank}_q(c)$ be the first rank of chunk $c$, or $\infty$ if it is absent from the returned list. A group's completion rank is the position of its last required member, and the question's earliest completion rank is the best of these alternatives:
\begin{equation}
\begin{aligned}
\rho_q(C)&=\max_{c\in C}\operatorname{rank}_q(c),\\
\rho_q^*&=\min_{C\in\mathcal{C}_q}\rho_q(C).
\end{aligned}
\end{equation}
At an explicit evaluation depth $D$, Complete-Evidence Reciprocal Rank and its macro-average are
\begin{align}
\mathrm{CE\_RR}@D(q)&=
\begin{cases}
1/\rho_q^*, & \rho_q^*\leq D,\\
0, & \text{otherwise},
\end{cases}\\
\mathrm{CE\_MRR}@D&=\frac{1}{|Q_{\mathrm{CE}}|}\sum_{q\in Q_{\mathrm{CE}}}\mathrm{CE\_RR}@D(q).
\end{align}
The reported depths are $D=20$ and $D=1000$. Chunks may arrive in any order; the last required member determines completion. For groups $\{\{A,B\},\{A,C\}\}$ and ranking $(A,D,C,B)$, ordinary reciprocal rank is 1, CE\_Recall@2 is 0, CE\_Recall@3 is 1, and CE\_RR@20 is $1/3$.

\subsection{BM25 configuration and reproducibility}
\label{sec:bm25-configuration}
\label{sec:reproducibility}
The BM25 baseline is implemented with Pyserini~\cite{lin2021pyserini}, which provides a Python interface to the Lucene-based Anserini retrieval toolkit. Only the \texttt{chunk} field is indexed, and the original question string is submitted as the query. Identifiers name the indexed documents; category labels, relevance judgments, and bibliographic metadata are excluded from the searchable text.

Parameters are fixed at $k_1=0.9$ and $b=0.4$, without tuning on the evaluation questions. The underlying Anserini English analyzer applies stopword removal and Porter stemming to the index and queries. Retrieval returns up to 1,000 matching chunks per question, retaining shorter lists when fewer match. All metrics use the same saved rankings and released relevance annotations.

The saved \texttt{run\_bm25\_top1000.trec} ranking can be rescored with \texttt{scripts/evaluate\_bm25.py} without rebuilding the index. The script writes per-question scores, overall and category averages, and group-completion outputs in \texttt{metric\_outputs}. The repository linked in Section~\ref{sec:data-availability} documents dependencies and commands for retrieval and scoring.

Comparisons should retain the corpus, question wording, released relevance labels, and reported cutoffs. Changes to those inputs should be identified alongside the method settings so that performance differences can be attributed to the appropriate experiment.

%% file: sections/09_results.tex
\section{Results}
\label{sec:results}
Table~\ref{tab:bm25-main} reports MRR@20 of 0.315, Recall@10 of 0.182, and nDCG@10 of 0.205 across all 325 questions. Inspection of the per-question scores shows no annotated hit in the first 10 results for 167 questions (51.4\%). The low aggregate coverage therefore accompanies many questions for which the short result list contains no gold passage.

\begin{table}[!htbp]
\caption{Overall BM25 results on Dh\={a}r\={a} Bench, macro-averaged over all 325 questions.}
\label{tab:bm25-main}
\centering
\begin{tabular}{lc}
\toprule
Metric & Macro score \\
\midrule
MRR@20 & 0.315 \\
Precision@10 & 0.115 \\
Recall@10 & 0.182 \\
F1@10 & 0.107 \\
nDCG@10 & 0.205 \\
Recall@20 & 0.229 \\
\bottomrule
\end{tabular}
\end{table}

The cutoff results in Table~\ref{tab:bm25-cutoff} show the coverage-precision tradeoff. From $k=3$ to $k=20$, Recall rises from 0.116 to 0.229 and Precision falls from 0.189 to 0.083. F1 peaks at $k=8$ (0.110) among the tested cutoffs; the larger lists recover more gold chunks but devote a smaller share of their positions to them.

\begin{table}[!htbp]
\caption{Overall macro BM25 performance on Dh\={a}r\={a} Bench for cutoff-dependent metrics.}
\label{tab:bm25-cutoff}
\centering
\begin{tabular}{rrrrr}
\toprule
$k$ & P@$k$ & R@$k$ & F1@$k$ & nDCG@$k$ \\
\midrule
3 & 0.189 & 0.116 & 0.107 & 0.225 \\
8 & 0.128 & 0.168 & 0.110 & 0.207 \\
10 & 0.115 & 0.182 & 0.107 & 0.205 \\
20 & 0.083 & 0.229 & 0.098 & 0.210 \\
\bottomrule
\end{tabular}
\end{table}

Category leaders differ by measure (Table~\ref{tab:bm25-cat}): T09 has the highest Recall@20 (0.391), T12 the highest nDCG@10 (0.331), T01 the highest MRR@20 (0.455), and T11 the highest Precision@3 (0.271). T08 has the lowest scores on all 4 measures. Section~\ref{sec:discussion} examines these differences alongside gold-set sizes.

\begin{table*}[!t]
\caption{Category-wise fixed BM25 results on Dh\={a}r\={a} Bench. Values are macro-averaged within each category.}
\label{tab:bm25-cat}
\centering
\begin{tabular}{lrrrrrrrr}
\toprule
Category & $n$ & MRR@20 & P@3 & P@10 & R@10 & F1@10 & nDCG@10 & R@20 \\
\midrule
T01 & 51 & 0.455 & 0.255 & 0.159 & 0.188 & 0.133 & 0.262 & 0.248 \\
T02 & 40 & 0.261 & 0.133 & 0.095 & 0.172 & 0.093 & 0.174 & 0.209 \\
T03 & 22 & 0.278 & 0.136 & 0.105 & 0.220 & 0.107 & 0.186 & 0.236 \\
T04 & 25 & 0.376 & 0.240 & 0.140 & 0.172 & 0.112 & 0.223 & 0.192 \\
T05 & 31 & 0.299 & 0.183 & 0.116 & 0.151 & 0.106 & 0.193 & 0.200 \\
T06 & 19 & 0.234 & 0.175 & 0.084 & 0.185 & 0.086 & 0.172 & 0.245 \\
T07 & 15 & 0.315 & 0.178 & 0.073 & 0.184 & 0.093 & 0.201 & 0.305 \\
T08 & 26 & 0.019 & 0.026 & 0.008 & 0.006 & 0.007 & 0.010 & 0.006 \\
T09 & 18 & 0.352 & 0.204 & 0.150 & 0.309 & 0.171 & 0.279 & 0.391 \\
T10 & 29 & 0.307 & 0.207 & 0.131 & 0.101 & 0.090 & 0.170 & 0.146 \\
T11 & 32 & 0.382 & 0.271 & 0.159 & 0.246 & 0.150 & 0.263 & 0.308 \\
T12 & 17 & 0.406 & 0.196 & 0.088 & 0.356 & 0.134 & 0.331 & 0.388 \\
\midrule
Overall & 325 & 0.315 & 0.189 & 0.115 & 0.182 & 0.107 & 0.205 & 0.229 \\
\bottomrule
\end{tabular}
\end{table*}

\begin{figure*}[!t]
\centering
\includegraphics[width=0.98\textwidth]{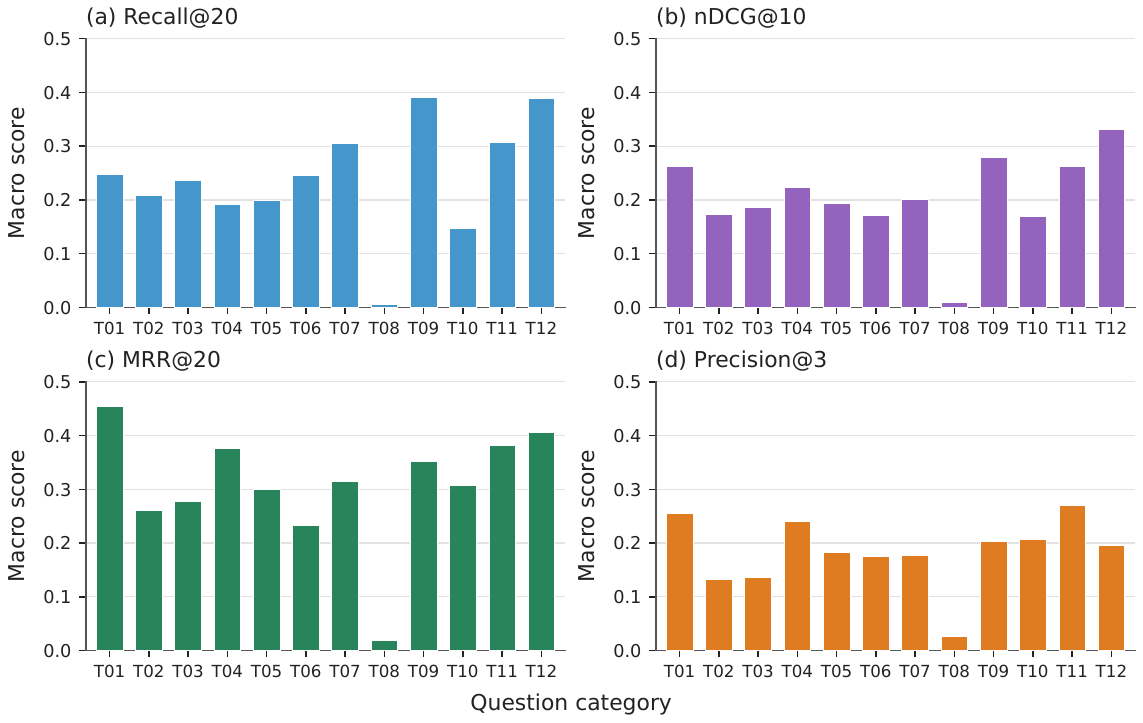}
\caption{Category-wise performance of the fixed BM25 baseline: (a) Recall@20, (b) nDCG@10, (c) MRR@20, and (d) Precision@3. Each bar is the macro-average within its question category; all panels use the same zero-based vertical scale.}
\label{fig:bm25-cat}
\Description{The 4 bar-chart panels compare categories T01 through T12. Blue bars show Recall at 20, purple bars nDCG at 10, green bars MRR at 20, and orange bars Precision at 3. T09 has the highest Recall at 20, 0.391; T12 the highest nDCG at 10, 0.331; T01 the highest MRR at 20, 0.455; and T11 the highest Precision at 3, 0.271. T08 has the lowest scores in all 4 panels. All vertical axes run from 0 to 0.5. Table~\ref{tab:bm25-cat} gives the complete results.}
\end{figure*}

Figure~\ref{fig:bm25-cat} plots these 4 category-wise comparisons on a common scale.

Table~\ref{tab:bm25-ce} reports complete-evidence results for the 19 T06 and 15 T07 questions. BM25 recovers at least one complete group within the top 10 results for 7 of these 34 questions (CE\_Recall@10 = 0.206), and within the top 20 for 10 (CE\_Recall@20 = 0.294). The pooled CE\_MRR@20 is 0.141, increasing to 0.144 at depth 1,000.

\begin{table*}[!t]
\caption{Complete-evidence BM25 results: CE\_Recall at cutoff $k$ and CE\_MRR at depth $D$. All annotated questions are evaluable; $n$ is the macro-average denominator.}
\label{tab:bm25-ce}
\centering
\begin{tabular}{lrrrrrrr}
\toprule
 & & \multicolumn{4}{c}{CE\_Recall@$k$} & \multicolumn{2}{c}{CE\_MRR@$D$} \\
\cmidrule(lr){3-6}\cmidrule(lr){7-8}
Scope & $n$ & $k=3$ & $k=8$ & $k=10$ & $k=20$ & $D=20$ & $D=1000$ \\
\midrule
T06 & 19 & 0.211 & 0.211 & 0.211 & 0.316 & 0.132 & 0.132 \\
T07 & 15 & 0.133 & 0.200 & 0.200 & 0.267 & 0.153 & 0.158 \\
T06+T07 & 34 & 0.176 & 0.206 & 0.206 & 0.294 & 0.141 & 0.144 \\
\bottomrule
\end{tabular}
\end{table*}

On the same T06 and T07 questions, MRR@20 is 0.234 and 0.315, while CE\_MRR@20 is 0.132 and 0.153. Requiring the last member of a valid group lowers the reciprocal-rank scores in both categories.

%% file: sections/09_discussion.tex
\section{Discussion}
\label{sec:discussion}
The results support different choices for passage lookup and evidence collection. For Question 1, ``What is Jatharagni?'', BM25 places a gold passage first, giving reciprocal rank 1. Yet it retrieves only 1 of the 21 gold chunks within 20 results, a recall of 4.8\%. A reader seeking one useful starting passage and a researcher assembling a broader set of references would experience this ranking differently. Reporting MRR alongside coverage makes that difference visible.

\subsection{Retrieval differences across question categories}
Verse retrieval is a priority for improvement on this collection. Of the 26 T08 questions, 25 have no gold hit within 20 results. Their median gold-set size is only 1.5 chunks (Table~\ref{tab:evidence-size}), so the poor coverage cannot be explained simply by sets that exceed the retrieval budget. Error inspection should examine how a question names or paraphrases its target, which scripts or transliterations occur in the indexed passage, and whether textual references are preserved. These checks can guide experiments with normalization, verse-specific indexing, or semantic matching. The scores identify the failure pattern; they do not determine which of these changes will resolve it.

The category rankings also resist a single ordering of task difficulty. Fact questions lead on first-hit rank, false-premise questions on Recall@20, and interpretive questions on Precision@3 (Figure~\ref{fig:bm25-cat}). T12's relatively high recall accompanies a median of 2 gold chunks, compared with 9 for T01. At the other extreme, the largest T10 set contains 2,549 chunks: even 20 gold hits would recover only 0.8\% of it. Researchers should therefore compare methods within the categories relevant to their application and examine annotation size when interpreting recall. A category average combines the requested task, the selected sources, and the breadth of its gold judgments.

\subsection{Choosing the retrieval budget}
Increasing the cutoff offers a measurable but limited tradeoff. Doubling the result list from 10 to 20 raises macro Recall by 4.7 percentage points, while Precision falls from 0.115 to 0.083 (Table~\ref{tab:bm25-cutoff}). The added passages may help evidence collection, but they also increase the material to inspect or pass to a model. F1 peaks at 8 results among the tested cutoffs; this balances the recorded judgments under that metric, without establishing the best budget for every application. An interface for browsing passages and a system assembling support for several subquestions can reasonably require different cutoffs. Chunk lengths also vary, so the text budget should be measured alongside the number of results.

\subsection{Recovering complete evidence groups}
Group recovery provides a separate decision criterion for composite and multi-hop queries. At 20 results, 24 of the 34 questions still lack a complete annotated group (Table~\ref{tab:bm25-ce}); expanding from 10 to 20 completes support for 3 additional questions. For these failures, the useful diagnostic is which required member is missing and where it appears in the saved ranking. Reranking can promote a member already in the candidate set; evidence absent from that set requires a change to candidate generation. CE\_Recall and CE\_MRR can then test whether the revised pipeline completes more groups and completes them earlier.

\subsection{Error analysis and method comparison}
The corpus records support inspection of these errors at the passage and source levels. A failed match may involve the retrieved wording, an edition-specific reference, OCR text, or a machine-translated commentary. OCR and translation accuracy were not independently evaluated, so their contribution to the observed errors requires source checking. Subsequent comparisons with dense or hybrid retrieval should retain the same corpus, questions, labels, and cutoffs. This provides a basis for deciding whether gains come from better candidate coverage, improved ordering, or more successful assembly of the annotated groups.

%% file: sections/10_conclusion.tex
\section{Conclusion}
Dh\={a}r\={a} Bench brings together an IKS corpus, 12 question categories, expert passage judgments, and source-linked records for shared retrieval evaluation. Its conjugate groups capture alternative combinations of passages supporting composite and multi-hop questions. The BM25 results expose two practical gaps: annotated passages are frequently absent from short result lists, especially for verse queries, and early access to one relevant chunk often leaves required support incomplete. Category-wise scores and group-completion measures make these gaps available for targeted comparison. The public corpus, rankings, and scoring scripts provide a common starting point for testing changes to candidate generation, ranking, and evidence assembly.

%% file: sections/acknowledgements.tex
\renewcommand{\acksname}{Acknowledgements}
\begin{acks}
An early proof of concept for this work was supported by the Department of Science and Technology (DST), Government of India, through its Science and Heritage Research Initiative (SHRI), under grant no.~DST/TDT/SHRI-34/2021. This work was supported by a Corporate Social Responsibility (CSR) grant from Perfios Software Solutions Private Limited to HTSR Institute. One of the joint first authors was supported through the Adani Indology Fellowship, funded by Adani Enterprises Limited.

We thank Adiyen Technologies Private Limited for assistance with data preparation.
\end{acks}